\RequirePackage{fix-cm}

\documentclass[twocolumn,epjc3]{svjour3}  
\pdfoutput=1 
\smartqed 
\RequirePackage{graphicx}
\usepackage{units}
\usepackage{amsmath}
\usepackage{subfigure}
\usepackage{lineno}
\usepackage[hyperindex,breaklinks]{hyperref} 
\newcommand{\Wtaunu}{\ensuremath{W\rightarrow\tau\nu}}
\newcommand{\ipb}{\ensuremath{\mathrm{pb}^{-1}}}
\newcommand{\Ptau}{\ensuremath{P_\tau}}
\newcommand{\Ztautau}{\ensuremath{Z\to\tau\tau}}
\newcommand{\Zmumu}{\ensuremath{Z\to\mu\mu}}
\newcommand{\Wmunu}{\ensuremath{W\to\mu\nu}}
\newcommand{\Wenu}{\ensuremath{W\to e\nu}}
\newcommand{\Wtaunulep}{\ensuremath{W\to\tau_{\mathrm{lep}}\nu}}
\newcommand{\GeV}{\mathrm{GeV}}
\newcommand{\TeV}{\mathrm{TeV}}

\newcommand{\taum}{\ensuremath{\tau^-}}
\newcommand{\taup}{\ensuremath{\tau^+}}
\newcommand{\ttbar}{\ensuremath{t\bar{t}}}
\newcommand{\pt}{\ensuremath{p_{\mathrm T}}}

\newcommand{\et}{\ensuremath{E_{\mathrm T}}}
\newcommand{\met}{\ensuremath{E_{\mathrm T}^{\mathrm{miss}}}}
\newcommand{\metsig}{\ensuremath{S_{E_{\mathrm T}^{\mathrm{miss}}}}}

\usepackage{preprintcover}  
\PreprintCoverPaperTitle{Measurement of $\tau$ Polarization in $\boldmath{\Wtaunu}$ Decays with the ATLAS Detector in $pp$ Collisions at $\sqrt{s}~=~7~\TeV$} 
\PreprintIdNumber{CERN-PH-EP-2012-075} 
\PreprintCoverAbstract{In this paper, a measurement of $\tau$ polarization in $\Wtaunu$ decays is presented. It is measured from the energies of the decay products in hadronic $\tau$ decays with a single 
final state charged particle. The data, corresponding to an integrated luminosity of 24~$\ipb$, were collected by the ATLAS experiment at the Large Hadron Collider in 2010. The measured value of the 
$\tau$~polarization is $\Ptau = -1.06\; \pm~0.04\;(stat)\; ^{+0.05}_{-0.07}\;(syst)$, in agreement with the Standard Model prediction, and is consistent with a physically allowed 95\% CL interval~$[-1,-0.91]$.
Measurements of $\tau$~polarization have not previously been made at hadron colliders.
}  
\PreprintJournalName{European Physical Journal C}

\begin{document}

\title{Measurement of $\tau$ Polarization in $\boldmath{\Wtaunu}$ Decays with the ATLAS Detector in $pp$ Collisions at $\sqrt{s}~=~7~\TeV$
}

\author{The ATLAS Collaboration}
\institute{CERN,1211 Geneva 23, Switzerland, email: atlas.publication@cern.ch} 
\thankstext{e1}{e-mail: atlas.publication@cern.ch}
\thankstext{add1}{CERN,1211 Geneva 23, Switzerland}
\date{}

\maketitle

\begin{abstract}
In this paper, a measurement of $\tau$ polarization in $\Wtaunu$ decays is presented. It is measured from the energies of the decay products in hadronic $\tau$ decays with a single final state charged particle. 
The data, corresponding to an integrated luminosity 
of 24~$\ipb$, were collected by the ATLAS experiment at the Large Hadron Collider in 2010. The measured value of the $\tau$~polarization is $\Ptau = -1.06\; \pm~0.04\;(stat)\; ^{+0.05}_{-0.07}\;(syst)$, in agreement with the Standard Model 
prediction, and is consistent with a physically allowed 95\% CL interval~$[-1,-0.91]$. 
Measurements of $\tau$~polarization have not previously been made at hadron
colliders.
\end{abstract}

\section{Introduction}
\label{sec:Intro}
The $\tau$~lepton plays an integral role in the physics program at the Large Hadron Collider (LHC) as a powerful probe in searches for new phenomena.
As the most massive lepton and a third generation particle, the $\tau$~lepton is particularly relevant in probing the nature of electroweak symmetry breaking.  
The branching fraction of the Standard Model (SM) Higgs boson to $\tau$~pairs is large in the low-mass region currently favored by experiment~\cite{HIGGSATLAS,HIGGSCMS}. In some regions of supersymmetry parameter space, decay chains with $\tau$~leptons provide discovery channels, for example at high values of $\tan \beta$ for the Minimal Supersymmetric Standard Model (MSSM) charged Higgs boson~\cite{MSSMTRY}. Due to the short-enough lifetime of $\tau$~leptons and their parity-violating decays, $\tau$~leptons are the only leptons whose spin information is preserved in the decay product kinematics recorded in the ATLAS detector. The $\Wtaunu$ coupling at low $W$ virtuality ($Q^2$), which governs the tau decay kinematics, is well known~\cite{PDG} while the helicity structure at $Q^2=m_\mathrm{W}^2$ has not been explicitly measured before. 

The $\tau$ polarization, $\Ptau$, is a measure of the asymmetry of the cross-section for left-handed and right-handed $\tau$~production, defined by
\begin{equation}
\Ptau = \frac{\sigma_R - \sigma_L}{\sigma_R + \sigma_L}
\label{eqn:ptau}
\end{equation}
for the production of $\taum$. While it is the $\tau$~helicity states that are experimentally accessible, the positive (negative) helicity states and right-handed (left-handed) chiral states coincide in the relativistic limit assumed here. CP invariance holds in $\tau$~decays in general~\cite{PDG}, and therefore the distributions for left-handed (right-handed) $\taup$ follow those of right-handed (left-handed) $\taum$. 
The value of $\Ptau$ provides insight into the Lorentz structure in the $\tau$~production mechanism. In particular, it is a measure of the degree of parity violation in the interaction. In $\Wtaunu$ decays, the $W^-$ is expected to couple exclusively to a left-handed $\taum$ and the $W^+$ to a right-handed $\taup$ corresponding to a $\tau$~polarization of $\Ptau=-1$. 
A parity-conserving decay results in a value of $\Ptau=0$. This is the case for the decay of a SM scalar Higgs boson to $\tau$~lepton pairs. On the other hand, an MSSM charged scalar Higgs boson couples to $\tau$~leptons leading to a prediction of $\Ptau=+1$.

The method outlined here for extracting the $\tau$~polarization is independent of the mode of $\tau$~production and can be applied to the characterization of new phenomena at the LHC. In particular, $\Ptau$ may be used as a discriminating variable in searches for new particles that decay to $\tau$~leptons and, in the event of such a discovery, could provide insight into the nature of the new particle's couplings.

\section{Tau Polarization Observables}
\label{sec:Theory}

Parity is maximally violated in the charged-current weak decays of $\tau$~leptons whereby the $\taum$ always couples to a left-handed $\tau$~neutrino, $\nu$. Due to angular momentum conservation, the angular distribution of the $\tau$~decay products depends strongly on the spin orientation of the $\tau$~lepton. The hadronic decay modes are particularly well suited to determine the $\tau$~spin orientation due to the fact that there is only one neutrino in the final state.

The angle $\theta$ between the $\tau$~direction of flight and hadronic decay products in the $\tau$~rest frame is the primary observable sensitive to $\tau$~polarization. The dependence of the angular distributions of the $\tau$~decay products on the $\tau$~polarization is discussed in detail in \linebreak Refs.~\cite{B235,C48}. In the relativistic limit, $E \gg m_\tau$, the angle~$\theta$ is related to the ratio of the energy of the hadronic decay products to the $\tau$~energy in the laboratory frame. The reconstruction of the $\tau$~energy in $\Wtaunu$ decays, however, is limited experimentally due to poor resolution arising from the multiple unobserved neutrinos in the final state. 
The $\tau$~lepton decay branching ratio to a single charged pion along with a neutral pion via an intermediate $\rho$ meson is about 25\%~\cite{PDG}.
For these decays an additional observable $\cos\psi$ is defined in the $\rho$ rest frame, where $\psi$ is the angle between the flight direction of the $\rho$ meson and the charged pion. 
This observable is related to the kinematics of the final state charged and neutral pions, which are experimentally accessible, as follows:
\begin{equation}
\cos \psi = \frac{m_{\rho}}{\sqrt{m_{\rho} ^2-4m_\pi^2}}\frac{E_{\pi^-}-E_{\pi^0}}{|{\mathbf p_{\pi^-}}+{\mathbf p_{\pi^0}}|},
\label{eqn:obs_thry}
\end{equation}
\noindent where the particle energies and momenta are measured in the laboratory frame. In $\tau\rightarrow\rho\nu$ decays, to conserve angular momentum, transversely polarized $\rho$ mesons are favored in left-handed $\tau$~decays while longitudinally polarized $\rho$ mesons are favored in right-handed $\tau$~decays. The transversely polarized $\rho$ decays to charged and neutral pions with comparable energies while the longitudinal $\rho$ results in an asymmetry in the energy sharing.

\section{The ATLAS Detector at the Large Hadron Collider}
\label{sec:Detector}

During 2010 operation, the LHC at the CERN laboratory provided proton-proton collisions with a center of mass energy
of 7 TeV. The ATLAS detector~\cite{ATLASDet} is a multi-purpose particle detector constructed with three primary detection systems layered radially as follows: a central inner tracking detector contained in a 2~T magnetic field providing charged particle position and momentum measurements, a calorimeter system for energy measurements of charged and neutral particles, and a muon spectrometer for measurements of positions and momenta of
muons in a magnetic field, provided by three large superconducting toroidal magnets, each consisting of eight coils.\footnote{ATLAS uses a right-handed coordinate system with its origin at the nominal interaction point (IP) in the center of the detector and the $z$-axis along the beam pipe. The $x$-axis points from the IP to the center of the LHC ring, and the $y$-axis points upwards. Cylindrical coordinates ($r,\phi$) are used in the transverse plane, $\phi$ being the azimuthal angle around the beam pipe. The pseudorapidity is defined in terms of the polar angle $\theta$ as $\eta=-\ln\tan\frac{\theta}{2}$. 
Transverse momentum and energy are defined as $p_{\mathrm T} = p \sin\theta$ and $\et = E \sin\theta$, respectively.} Measurements in the inner tracking detector are performed with silicon detectors in the pseudorapidity range $|\eta|<2.5$ followed radially by a straw-tube tracking detector in the range $|\eta|<2.0$. 
The calorimeter consists of several sub-systems, providing high resolution reconstruction of the topology of particle showers for $|\eta|<4.9$. The electromagnetic calorimeter consists of barrel ($|\eta|<1.5$), 
two endcap ($1.4<|\eta|<3.2$), and forward ($3.1<|\eta|<4.9$) components, each of which utilizes liquid argon as the active material. The hadronic calorimeter consists of a scintillating tile calorimeter ($|\eta|<1.7$), two liquid argon end-cap and forward calorimeters ($1.5<|\eta|<4.9$). Events are selected with a three-tiered trigger system, the first of which is a hardware-based trigger using reduced granularity information from the calorimeter and muon systems.
The subsequent trigger levels are software based, and have access to the full event readout information. They make use of algorithms similar to those employed in the offline reconstruction. 

\section{Data and Simulation Samples}
\label{sec:Samples}

During the data-taking periods considered for this measurement, the maximum instantaneous luminosity \linebreak was $2.1\times 10^{32}~\mathrm{cm}^{-2}\mathrm{s}^{-1}$
corresponding to an average of 3.8 interactions per bunch crossing. The data were collected using a trigger designed to select events with a hadronically decaying $\tau$~lepton with transverse momentum of the visible decay 
products $\pt~>~16~\GeV$ along with missing transverse momentum $\met~>~22~\GeV$~\cite{TAUTRIG}. 
Maximum efficiency for this trigger is only achieved for $\pt~>~30~\GeV$.
With the larger instantaneous luminosity in 2011, an unprescaled trigger with comparable thresholds was not feasible and therefore the present analysis uses only data recorded in 2010. The data-taking periods considered are those for which all detector subsystems were operational and for which the trigger was not prescaled. The data used in this analysis were triggered by applying requirements to $\tau$~candidates that were less stringent than those required for offline $\tau$~identification. 
This latter requirement restricts the integrated luminosity of the data sample from 34~$\ipb$ to 24~$\ipb$. 
An additional sample, with an integrated luminosity of 5.6~$\ipb$, collected with very loose trigger requirements applied to the hadronically decaying $\tau$,~is used to evaluate the background contribution that results from the production of multijet events. 

Two $\Wtaunu$ Monte Carlo (MC) simulation samples were produced with a center of mass energy of \linebreak $\sqrt{s}~=~7$~TeV and with $\tau$~leptons decaying hadronically. 
The events were generated using HERWIG++~\cite{HERWIGPP} with the modified leading order (LO) parton distribution function (PDF)
MRSTLO*~\cite{PDF} and with $\tau$~leptons forced to decay by HERWIG++ as left-handed $\tau$~leptons in 
one sample and as right-handed $\tau$~leptons in the other. Simulation of the underlying event was improved by introducing a color reconnection model~\cite{UEHER}. In both samples, the $W$ boson production, $q\bar{q}\rightarrow W$, was simulated assuming Standard Model couplings of the $W$ boson to quarks.

The electroweak processes that may contribute to the background are leptonic $W$ boson decays to muons and electrons, $\Wtaunu$ decays in which the $\tau$ decays leptonically, $Z$ boson decays to lepton pairs, and top-quark pair ($\ttbar$) production. All Monte Carlo samples used to evaluate these backgrounds were generated with \linebreak PYTHIA~\cite{PYTHIA} with MRSTLO* PDFs except for the $\ttbar$ sample, which was produced with the MC@NLO~\cite{MCATNLO} generator and CTEQ6.6 PDFs, where parton showers and hadronization were simulated with HERWIG~\cite{HERWIG} and the underlying event with JIMMY~\cite{JIMMY}. The $\tau$~decays in these processes were modeled with \linebreak TAUOLA~\cite{TAUOLA1,TAUOLA2,TAUOLA3}, and QED radiation of photons was modeled with PHOTOS~\cite{PHOTOS}. 

All Monte Carlo samples were generated with multiple proton-proton interactions per bunch crossing and passed through the ATLAS detector simulation~\cite{GEANT4A} based on the GEANT4~\cite{GEANT4} package with the ATLAS MC10 settings~\cite{ATLASMC10}. 
The simulated events were \linebreak re-weighted so that the distribution of the number of reconstructed vertices matched that observed in data.

\section{Physics Object Reconstruction and Identification}
\label{sec:Obj}

The methods for reconstruction and identification of physics objects used in this analysis are identical to those used for the measurement 
of the \Wtaunu~cross-section documented in Ref.~\cite{WtauXS}.
 
Electron candidates are reconstructed from a cluster in the electromagnetic calorimeter that is matched to an inner detector track~\cite{ELERECO}.
Discrimination against backgrounds from jets, heavy quarks, and photon conversions is achieved using calorimeter shower shape and track quality requirements.
Muon candidates are reconstructed from tracks in the inner detector and muon spectrometer.

Jets are reconstructed using the anti-$k_t$ algorithm with a radius parameter $R~=~0.4$~\cite{ANTIKT}. 
All jet energies are calibrated with a $\pt$- and $\eta$-dependent energy scale~\cite{JES}, including corrections for losses in dead material and outside the jet cone~\cite{JCAL}. Jets seed the reconstruction of $\tau$ candidates where, in this analysis, ``$\tau$~candidate'' refers to a reconstructed object that resembles a hadronically decaying $\tau$~lepton. The $\tau$~direction is defined to be along the seed jet axis and the reconstructed $\tau$~energy is calculated from topological clusters~\cite{TOPO} in the calorimeter. The $\tau$~energy is subsequently calibrated with the $\tau$~energy scale, derived from Monte Carlo simulations and applied to the sum of energies of the cells that comprise the clusters of the seed jet~\cite{BDT1}. 
The $\tau$~candidate's pseudorapidity, $\eta$, and transverse momentum, $p_{\mathrm T}$, therefore refer to the visible decay products observed
in the detector.
In the calculation of the transverse momentum, the mass of the $\tau$~candidates is neglected since $E_{\tau}\gg m_{\tau}$, and therefore $\pt=E\sin\theta$. The $\tau$~candidates are required to have $\pt$ between 20~$\GeV$ and 60$~\GeV$ and to lie within $|\eta|<2.5$, and not in the calorimeter barrel-endcap transition region defined by 1.3~$<|\eta|<$~1.7. 

Hadronically decaying $\tau$~leptons, in contrast to the large multijet background, are characterized by low track multiplicity as well as narrow, isolated showers in the calorimeter. The identification of $\tau$~candidates is based on eight variables combined in a boosted decision tree algorithm~\cite{BDT1}. The most powerful discriminating variable is the electromagnetic radius, defined as the energy-weighted shower radius in $\eta-\phi$ space, calculated in the first three layers of the electromagnetic calorimeter~\cite{WtauXS}. In addition, a dedicated algorithm based on calorimeter variables is used to reject electrons that otherwise fake $\tau$~candidates.

The reconstruction of the missing transverse \linebreak momentum~($\met$) and total transverse energy~($\sum{\et}$) in the event is based on electromagnetic-scale energy deposits in calorimeter cells inside topological clusters~\cite{ETMISS}. The cluster energies are corrected for hadronic response, dead material and out-of-cluster losses. If a muon is present in the event, the $\met$ is corrected by including the muon $\pt$ in the total transverse energy. Based on these quantities, $\met$ significance, $\metsig$, is defined as:
\begin{equation}
\metsig~=~\frac{\met}{\sigma(\met)},
\end{equation}
\noindent where the $\met$~resolution has been approximated as $\sigma(\met)~=~0.5\sqrt{\GeV}~\sqrt{\sum{\et}}$.

\section{Event Selection and Background Estimation}
\label{sec:Selection}

Upon passing the trigger requirement, signal events are selected with criteria identical to those implemented in the measurement of the $\Wtaunu$ cross-section~\cite{WtauXS} with an additional requirement that the identified $\tau$~candidate has a single track reconstructed in the inner detector.
The main selection requirements, which are described in detail in the reference above, comprise an identified single-track $\tau$~candidate and
a minimum $\met$ of 30$~\GeV$, approximating the vector sum of the transverse energy of the final state neutrinos, $(\sum p^{\nu}$)$_\mathrm{T}$. 
In order to ensure a uniform $\met$ resolution, events are rejected if a jet with $\pt>20~\GeV$ and within the calorimeter barrel-endcap transition region is found. 
To further suppress the backgrounds, a minimum separation $|\Delta\phi(\mathrm{jet},\met)|>$~0.5 radians is required for jets with $\pt>20~\GeV$, since a large reconstructed $\met$ collinear with a jet is most likely due to misreconstruction of the jet energy. Multijet events are effectively suppressed with the requirement that $\metsig > 6$. Finally, events with an identified electron or muon with $\pt~>~15~\GeV$ are vetoed to suppress the background contributions from leptonic decays of $W$ and $Z$ bosons. 
The phase space for this measurement is therefore restricted by the conditions specified in Table~\ref{tab:truespace}.

\begin{table}[h]
\begin{center}
\caption{Definition of the phase space for the measurement of $\tau$~polarization in terms of the true $\tau$~visible decay product and neutrino kinematic variables~\cite{WtauXS}.}
\label{tab:truespace}
\begin{tabular}{c}
\hline\noalign{\smallskip}
Acceptance Region\\
\hline\noalign{\smallskip}
20~GeV$~<~p_{\mathrm{T}}^{\tau, \mathrm{vis}}~<~$60~GeV\\
$|\eta^{\tau,\mathrm{vis}}|~<~$2.5, excluding 1.3$~<~|\eta^{\tau,\mathrm{vis}}|~<~$1.7\\
($\sum p^{\nu}$)$_\mathrm{T}~>~$30~GeV\\
$|\Delta\phi(p^{\tau, \mathrm{vis}},\sum p^{\nu})|~>~$0.5~rad\\
\noalign{\smallskip}\hline
\end{tabular}
\end{center}
\end{table}

\begin{table}[h]
\begin{center}
\caption{Number of events passing the full event selection for data, the Monte Carlo electroweak background estimate (which is independent of the assumed $\tau$~polarization), the Monte Carlo $\Wtaunu$ signal for left-handed and right-handed polarization, and the multijet background estimated from data. The multijet background estimates are corrected for signal contamination and therefore depend on the $\tau$~polarization. The data should be compared to the sum of the appropriate $\Wtaunu$ signal entry plus the electroweak and multijet background estimates. The errors denote the statistical uncertainty from the Monte Carlo samples.}
\label{tab:cutflow}
\begin{tabular}{ll}
\hline\noalign{\smallskip}
Sample & Number of Events \\ 
\noalign{\smallskip}\hline\noalign{\smallskip}
Data &1136 \\
Electroweak Background &138~$\pm~$4 \\ \hline
\multicolumn{2}{c}{Left-Handed Signal}\\ 
$W\rightarrow \tau_L\nu$ &1002$~\pm~$16 \\ 
Multijet Background &69 $~\pm~$ 6 \\ \hline
\multicolumn{2}{c}{Right-Handed Signal}\\ 
 $W\rightarrow \tau_R\nu$ &1523$~\pm~$22 \\ 
Multijet Background &79$~\pm~$4 \\
\noalign{\smallskip}\hline
\end{tabular}
\end{center}
\end{table}

\begin{figure}[ht]
\subfigure [Track \pt]{\includegraphics[width = \columnwidth]{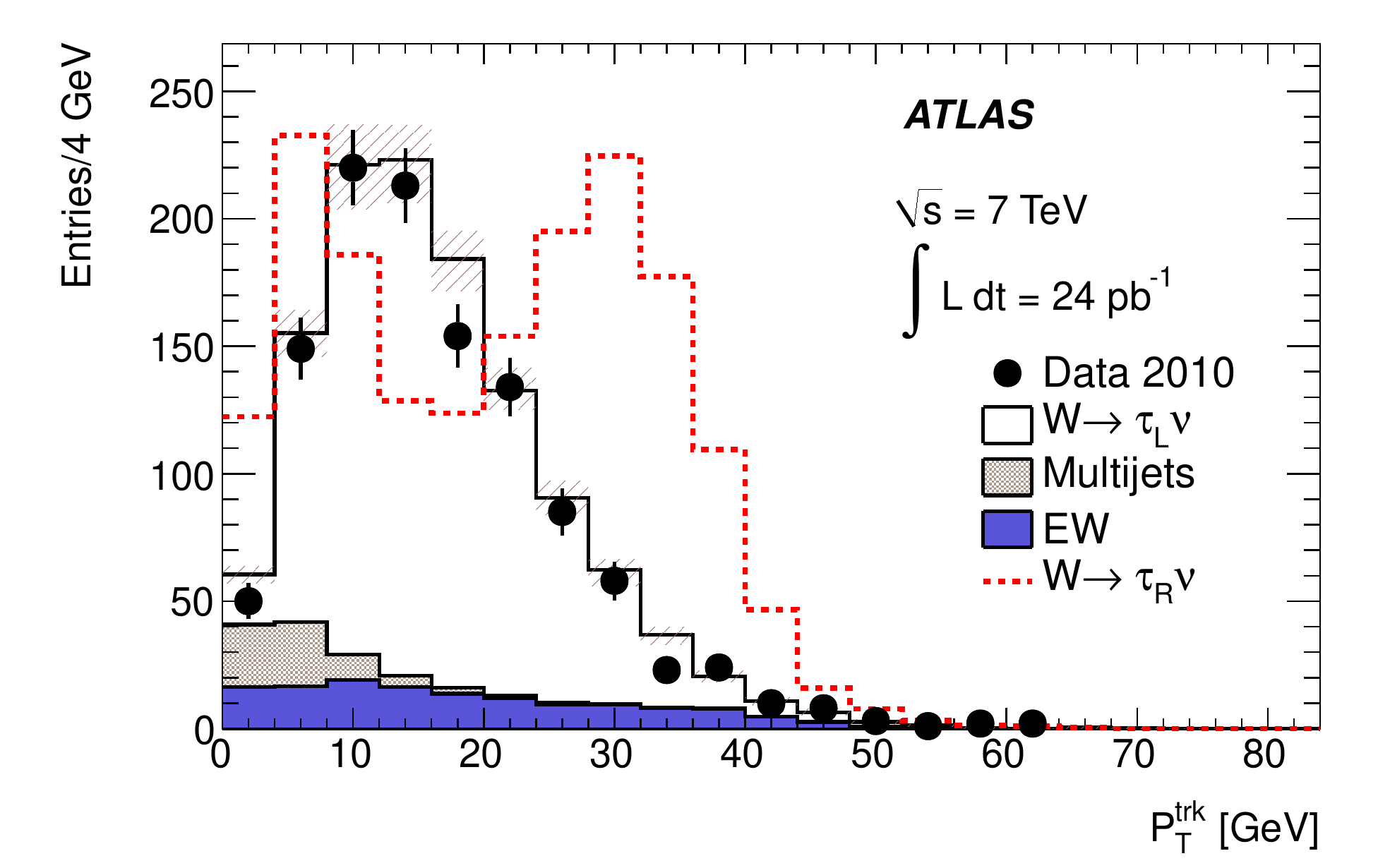}}

\subfigure[Tau \pt]{\includegraphics[width = \columnwidth]{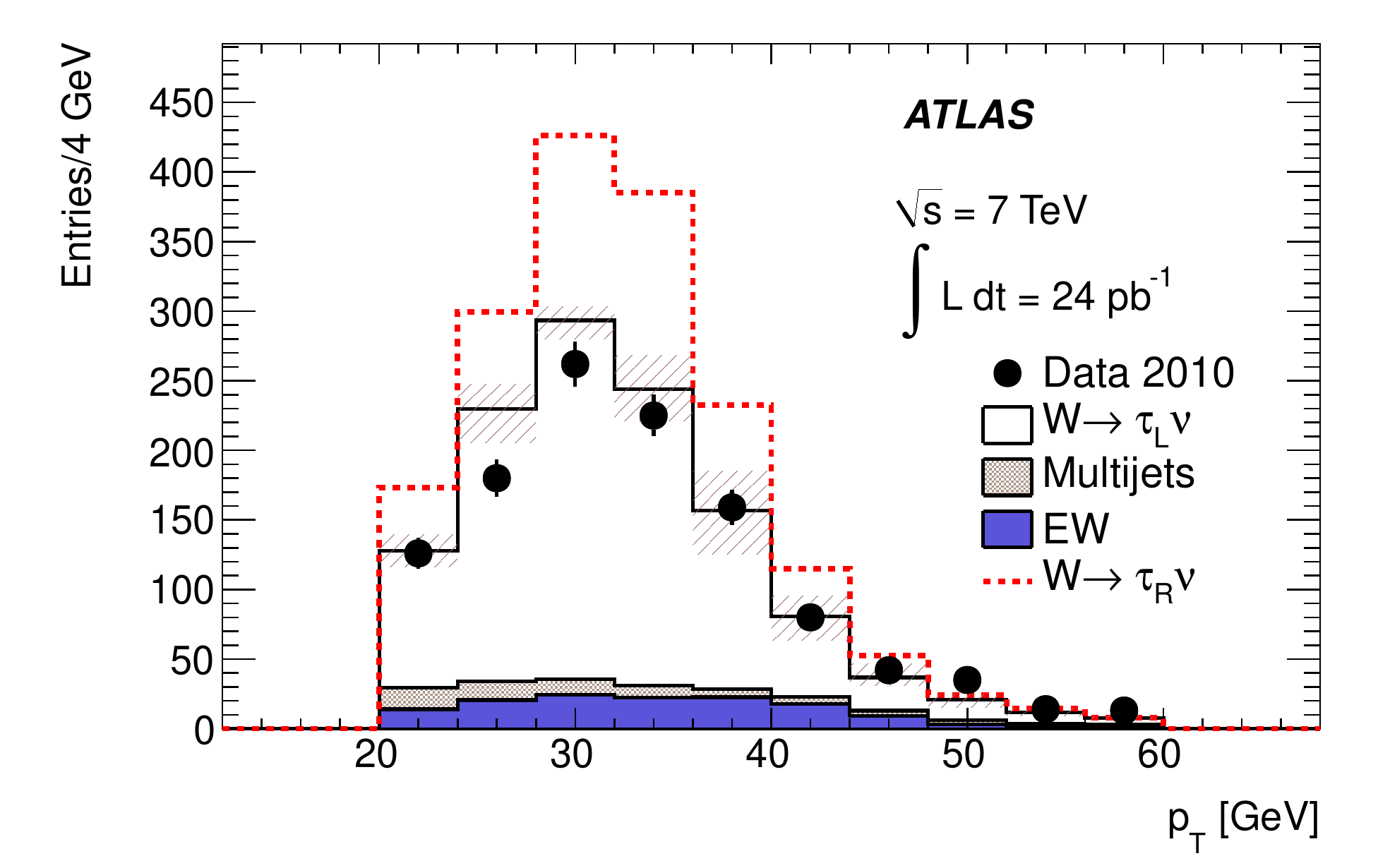}}

\subfigure[EM radius]{\includegraphics[width = \columnwidth]{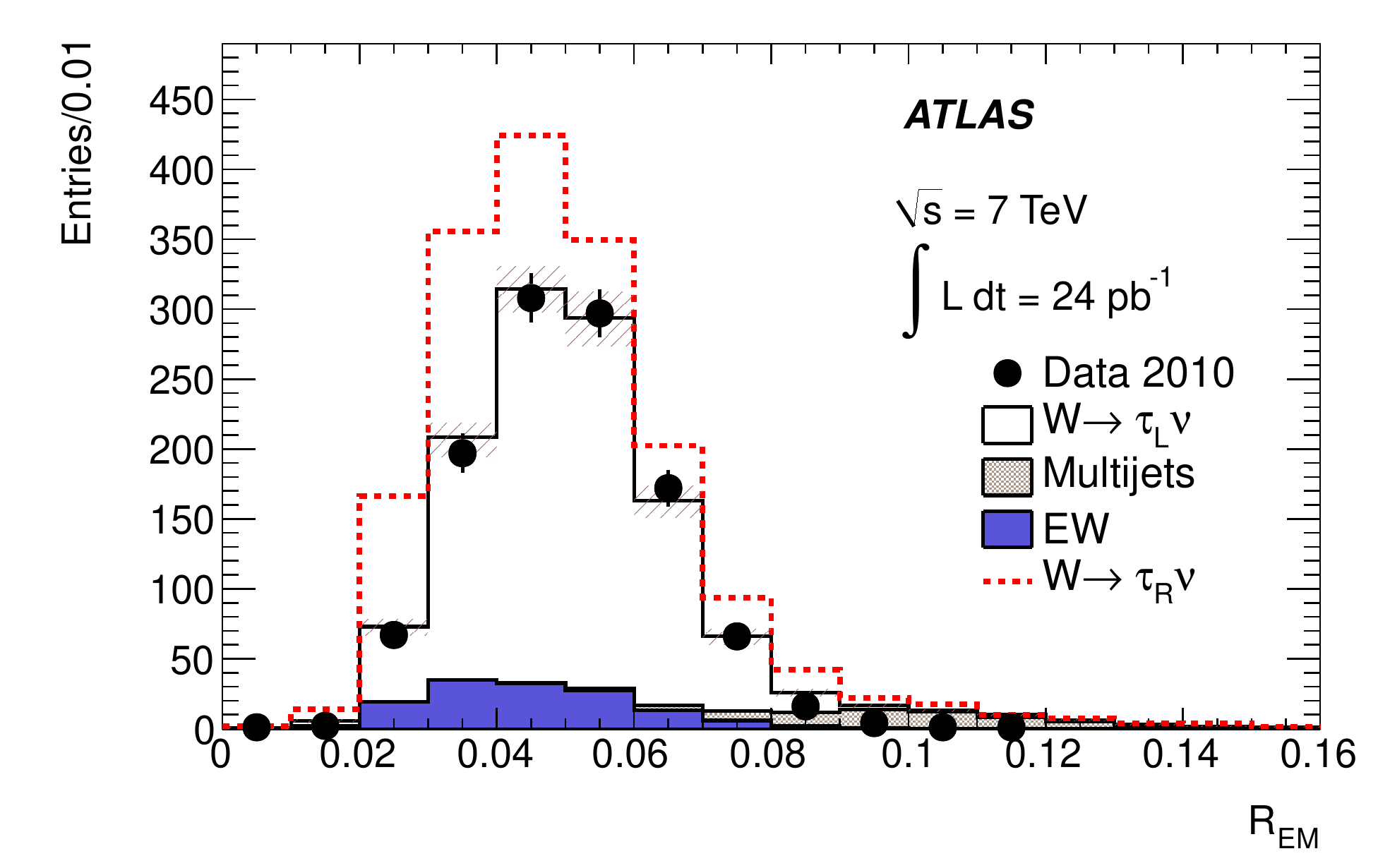}}

\caption{Kinematic distributions for $\tau$~candidates. The combined statistical and energy scale systematic uncertainties are overlaid on the stacked left-handed signal and background distributions. The stacked right-handed $\Wtaunu$ simulated signal and background distribution is also shown.}
\label{fig:taukinematics}
\end{figure}

\begin{figure}[ht]
\subfigure[Missing transverse momentum]{\includegraphics[width = \columnwidth]{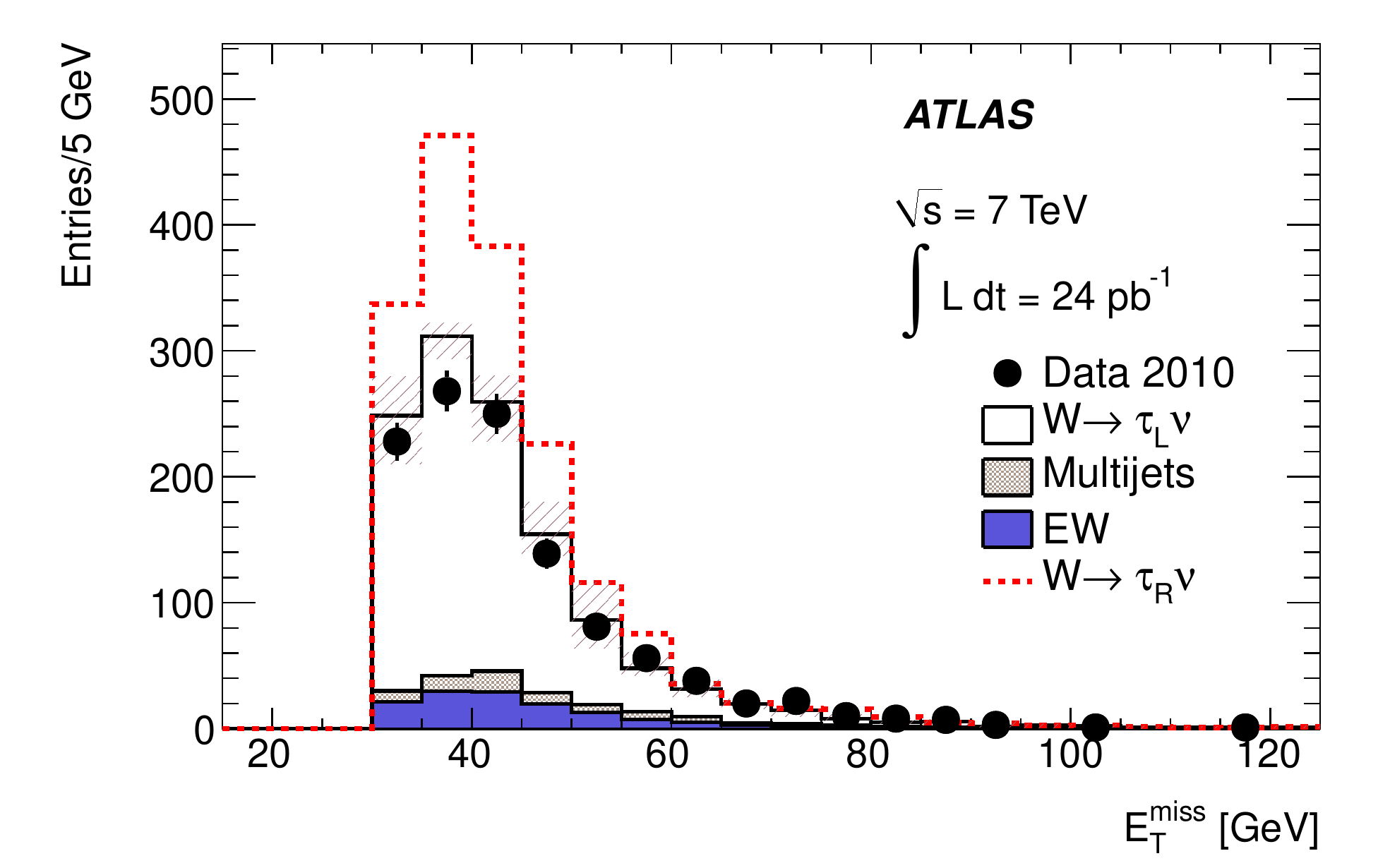}}

\subfigure[Missing transverse momentum significance]{\includegraphics[width =\columnwidth]{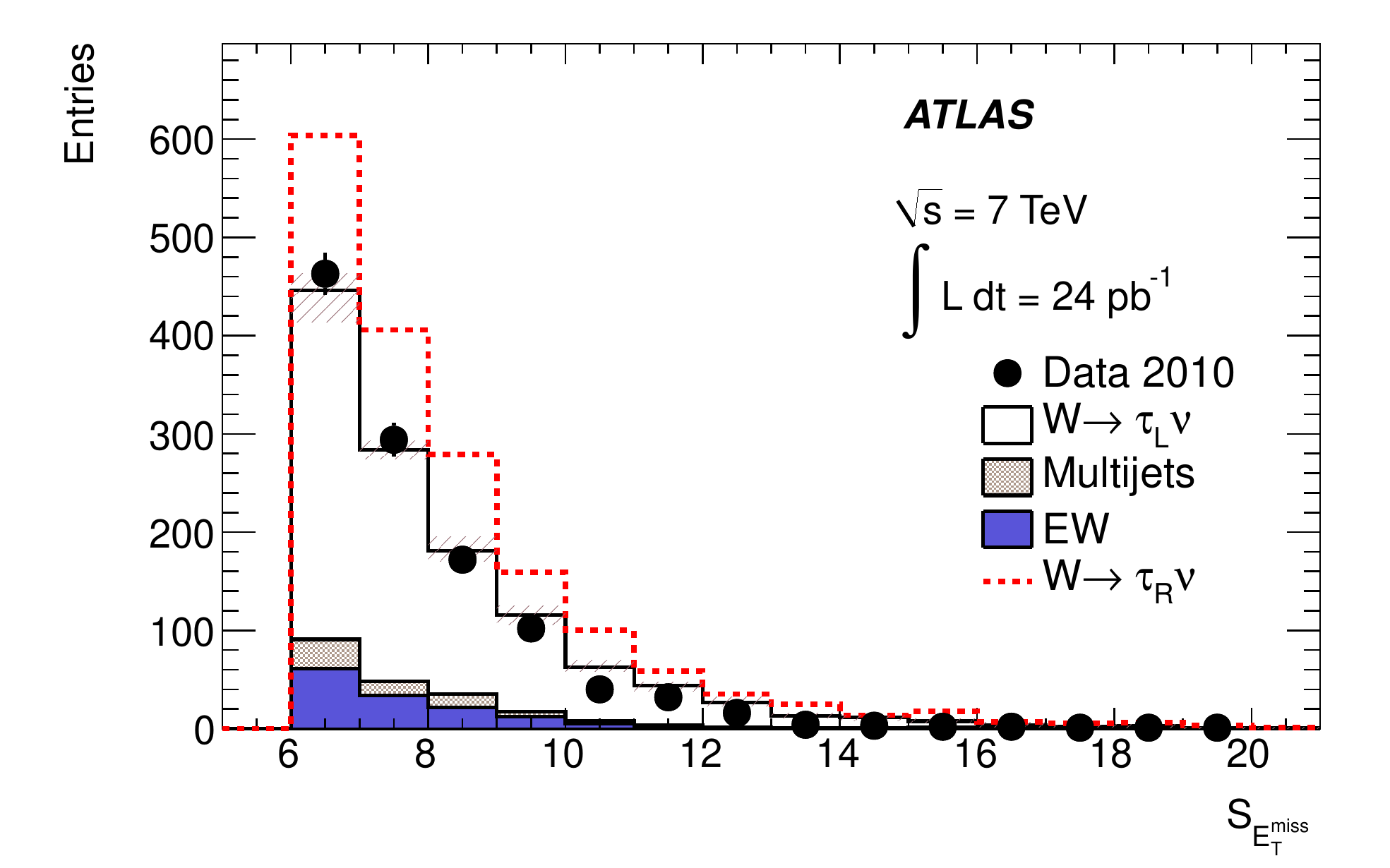}}

\caption{Event kinematic distributions. The combined statistical and energy scale systematic uncertainties are overlaid on the stacked left-handed signal and background distributions. The stacked right-handed $\Wtaunu$ simulated signal and background distribution is also shown. }
\label{fig:evkinematics}
\end{figure}

The electroweak backgrounds are estimated by applying the event selection to fully simulated samples of the relevant background processes and normalizing the contributions to the integrated luminosity of the data sample with the corresponding cross-sections measured by ATLAS~\cite{WZxsec,ttbar}. The multijet background is estimated from data following the method employed in the $\Wtaunu$ cross-section measurement~\cite{WtauXS}. The estimate is performed with a data sample collected with a looser combined $\tau$~and $\met{}$ trigger together with the signal sample. Three independent control regions are identified in addition to the signal region, separated by $\tau$~identification requirements and the value of the $\met{}$ significance. The shapes of kinematic distributions are taken from the region rich in multijets defined by a looser $\tau$~identification and $\metsig < 4.5$. The multijet background normalization in the signal region is estimated through a comparison of the number of events passing the full selection in the four statistically independent regions, and is corrected for contamination from electroweak events in the control regions, which ranges from approximately 1\% to 35\%.

The expected numbers of signal and background events passing the event selection are summarized in Table~\ref{tab:cutflow}, along with the number of selected events in data. Due to signal contamination in the control regions, the overall multijet background normalization depends on the $\tau$~polarization. Figure~\ref{fig:taukinematics} compares the track $\pt$, $\tau$~lepton~$\pt$, and the electromagnetic radius of the identified $\tau$~leptons in data and the left-handed (right-handed) $\Wtaunu$ signal plus the estimated electroweak and multijet backgrounds after the full event selection. Figure~\ref{fig:evkinematics} shows the same comparison for the $\met{}$ and $\met{}$ significance.

\section{Tau Polarization Observable in Data}

\begin{figure}[b]
\subfigure[Left-handed]{\includegraphics[width= \columnwidth]{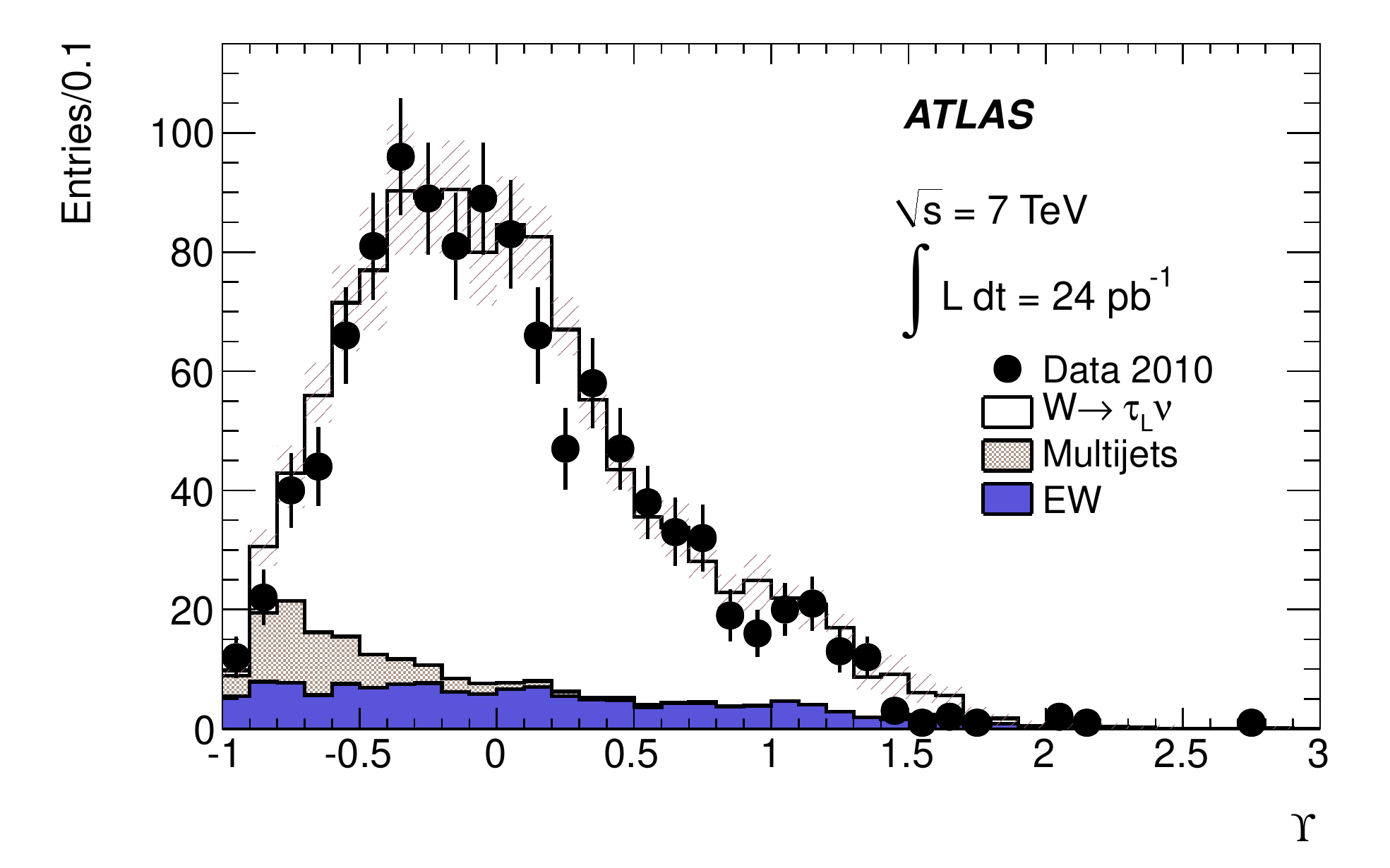}}

\subfigure[Right-handed]{\includegraphics[width= \columnwidth]{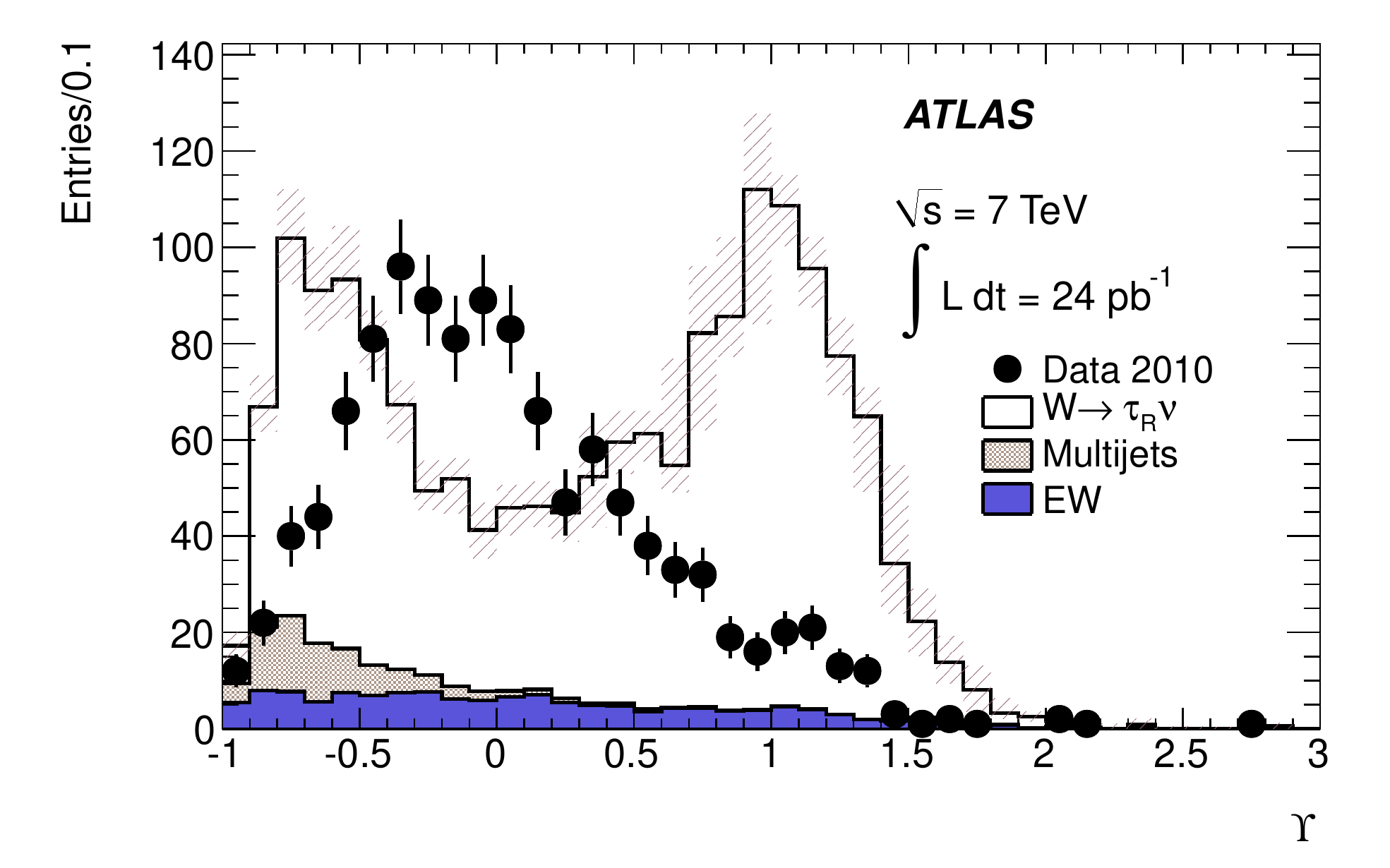}}
\caption{Charged asymmetry distributions.
The combined statistical and energy scale systematic uncertainties are overlaid on the stacked signal and background distributions.
}
\label{fig:ChargedAsymm}       
\end{figure}

From Equation~(\ref{eqn:obs_thry}), an observable, referred to as the ``charged asymmetry,'' is derived.
It measures the energy sharing of the charged and neutral pions in the $\tau$~decay relative to the visible
momentum of the $\tau$~lepton.  
Experimentally, the energy associated with the \linebreak charged pion is given by the transverse momentum of the single track associated with the $\tau$~candidate. The energy ascribed to the neutral pion(s) is calculated as the difference between the $\tau$~lepton $\pt$ measured in the calorimeter and the track~$\pt$ of the $\tau$~candidate.
The charged asymmetry $\Upsilon$ is calculated as follows:
\begin{equation}
\frac{\et^{\pi^-}-\et^{\pi^0}}{\pt} \approx 2\frac{p_{\mathrm T}^{\mathrm{trk}}}{\pt}-1 = \Upsilon .
\end{equation}
Although the observable in Equation~(\ref{eqn:obs_thry}) is defined for $\tau\to\rho\nu$ decays, in this analysis the charged asymmetry is measured in all of the decay modes to a single charged meson inclusively, which comprise roughly 50\% of all $\tau$~decays. After the event selection, the $\tau$ decay modes with a single final state neutral pion account for 60\% and of the left-handed and 53\% of the right-handed simulated samples. Figure~\ref{fig:ChargedAsymm} shows the distribution of the charged asymmetry observable in the left-handed and right-handed Monte Carlo signal samples plus the estimated electroweak and multijet backgrounds along with the observed distribution in data. These distributions reveal the tendency of the left-handed $\tau$~leptons to decay to charged and neutral pions with similar energies whereas the corresponding energy sharing is asymmetric in the right-handed $\tau$~decays. Moreover, these distributions illustrate the power of the charged asymmetry observable in distinguishing between left-handed and right-handed $\tau$~leptons. 

\begin{table*}[ht]
\begin{center}
\caption{Sensitivity of the charged asymmetry observable at various stages in the simulation process. $\Ptau$ denotes the assumed polarization. }
\label{tab:sensitivity}
\begin{tabular}{llll}
\hline\noalign{\smallskip}
\small{Stage of Simulation} & $\Ptau$~=~--1&$\Ptau$~=~0&$\Ptau$~=~+1\\ \hline\noalign{\smallskip}
\small{Generator Level, No Selection} &0.32 &0.25 &0.26 \\
\small{Generator Level, $p_{\mathrm{T}}^{\tau, \mathrm{vis}}~>~20~\GeV$,} \small{$|\eta^{\tau, \mathrm{vis}}|~<~2.5$, ($\sum p^{\nu}$)$_\mathrm{T}~>~$30~GeV} & 0.57 &0.45 &0.53 \\
\small{Reconstruction and} \small{Full Event Selection} &0.46&0.37&0.40\\
\noalign{\smallskip}\hline
\end{tabular}
\end{center}
\end{table*}
A measure of the analyzing power of the observable is provided by the sensitivity $S$, defined as:
\begin{equation}
S = \frac{1}{\sqrt{N}\sigma} = \sqrt{\int \frac{g^2(\Upsilon)}{f(\Upsilon)+\Ptau g(\Upsilon)}d\Upsilon},
\end{equation}
\noindent where $\sigma$ is the relative statistical error expected for a sample of $N$ events and where
$f(\Upsilon)$ and $g(\Upsilon)$ are functions that satisfy
$W(\Upsilon) = f(\Upsilon) + \Ptau g(\Upsilon)$ for the decay distribution $W(\Upsilon)$~\cite{B306}.
The sensitivity of the charged asymmetry $\Upsilon$ in a measurement of $\tau$~polarization is evaluated after reconstruction and the full event selection. 
Table~\ref{tab:sensitivity} summarizes the resulting sensitivities for assumed $\tau$~polarization values of~$-1$, 0, and +1, with calculations at the generator level included for reference. The event selection improves the sensitivity of $\Upsilon$ by suppressing events from $\tau\to\pi\nu$ decays, which exhibit softer spectra of both $\met$ and $\tau$~lepton~$\pt$ than those of the $\tau\to\rho\nu$ channel. This suppression is enhanced for left-handed $\tau$ leptons from $\tau\to\pi\nu$ decays, which account for 7\% of the simulated signal compared to 27\% in the simulated right-handed signal, after the full event selection. Detector effects lead to a loss in sensitivity of approximately 20\%.
\section{Extracting the $\tau$~Polarization}
\label{sec:Method}
The $\tau$ polarization is measured by fitting the observed charged asymmetry distribution of single-track hadronically decaying $\tau$~candidates satisfying the full \linebreak $\Wtaunu$ event selection to a linear combination of templates prepared with the left-handed and right-handed Monte Carlo samples. The right-handed (left-handed) template consists of the charged asymmetry distribution in the right-handed (left-handed) signal in addition to the estimated contributions from the electroweak and multijet backgrounds. 
The electroweak background distributions are effectively independent of the $\tau$~polarization in $\Wtaunu$ decays and are therefore common to both templates. Since the estimated multijet background depends on the $\tau$~polarization, rather than fixing the multijet normalization in the templates, the normalization is included as a parameter in the fit. 

The $\tau$~polarization, $\Ptau$, is extracted from the fit by maximizing a binned log-likelihood function. The likelihood per bin, $\mathcal{L}[i]$, is constructed as the product of Poisson terms as follows:
\begin{equation}
\label{eqn:LLH}
\begin{split}
\mathcal{L}[i] &~= \frac{e^{-T_i}(T_i)^{N_i}}{N_i!} \times \\
& \prod_{k=\mathrm{L,R}} \frac{e^{-s_i^k}(s_i^k)^{S_i^k}}{S_i^k!} \cdot \prod_{j} \frac{e^{-b_i^j}(b_i^j)^{B_i^j}}{B_i^j!}\cdot \frac{e^{-q_i}(q_i)^{Q_i}}{Q_i!},
\end{split}
\end{equation}
where $j$ labels the electroweak background processes $j$=\{$\Ztautau$,$\Zmumu$,$\Wenu$,$\Wmunu$,$\Wtaunulep$,$\ttbar$\}. The first factor in Equation~(\ref{eqn:LLH}) describes the probability to observe $N_i$ events in data given an expected value of $T_i$, the linear combination of template contributions per bin, as prescribed in Equation~(\ref{eqn:temp}). The remaining factors are included in the likelihood to account for the finite sample sizes and give the probabilities to observe the actual event counts, $S_i^{\mathrm L}$ ($S_i^{\mathrm R}$), $B_i^j$, and $Q_i$, for given expected values $s_i^{\mathrm L}$ ($s_i^{\mathrm R}$), $b_i^j$, and $q_i$, in the left-handed (right-handed), $j^{th}$ background, and multijet samples, respectively. Each of the $S_i^{\mathrm L}$ ($S_i^{\mathrm R}$), $B_i^j$, and $Q_i$ is taken without scaling to the integrated luminosity in data.

The assignment of the linear combination of the contributions per template for each bin, $T_i$, is given by
\begin{equation}
\label{eqn:temp}
\begin{split}
T_i(&N_{\mathrm{MC}},\Ptau,N_{\mathrm{MJ}}) =\\
&N_{\mathrm{MC}}\cdot \Big[\Big(\frac{1-P_{\tau}}{2}\Big)s_i^L \mu_{s^L} + \Big(\frac{1+P_{\tau}}{2}\Big)s_i^R \mu_{s^R} \Big] + \\
&N_{\mathrm{MC}}\cdot \Big[ \sum_j b_i^j \mu_{b^j} \Big] +N_{\mathrm{MJ}}\cdot q'_i.   
\end{split}
\end{equation}
The left-handed and right-handed $\Wtaunu$ signal components are weighted with the parameter $\Ptau$, which is used to extract the value of the $\tau$~polarization. The signal and electroweak background Monte Carlo contributions are normalized relative to each other according to their SM cross-sections with the factors $\mu_{s^{L}}$, $\mu_{s^{R}}$, and $\mu_{b^{j}}$. The overall normalization of the contributions from the $\Wtaunu$ signals and electroweak background processes is fitted with a single parameter, $N_{\mathrm{MC}}$, common across bins $i$, which accounts for the potential disagreement between the number of events predicted in Monte Carlo and that observed in data.  The multijet background estimation is similarly normalized with a separate fitted parameter, $N_{\mathrm{MJ}}$, common across bins $i$. Furthermore, the multijet contribution is explicitly corrected for the contamination of signal and electroweak background events as follows:
\begin{equation}
\label{eqn:tempqcd}
\begin{split}
q'_i &=~ q_i - \\ 
&N_{\mathrm{MC}}\Big[n_{i\mathrm{,MJ}}^{\mathrm{EW}}-\Big(\frac{1-\Ptau}{2}\Big) s^{L}_{i\mathrm{,MJ}} - \Big(\frac{1+\Ptau}{2}\Big)s^{R}_{i\mathrm{,MJ}}\Big],
\end{split}
\end{equation}
where $s^{\mathrm L}_{i,\mathrm{MJ}}$ ($s^{\mathrm R}_{i,\mathrm{MJ}}$) and $n_{i, \mathrm{MJ}}^{\mathrm{EW}}$ are the number of left-handed (right-handed) signal and electroweak \linebreak background events per bin $i$ in the multijet-rich control sample, scaled to the integrated luminosity in data.

\begin{figure}
\centering
  \includegraphics[width=\columnwidth]{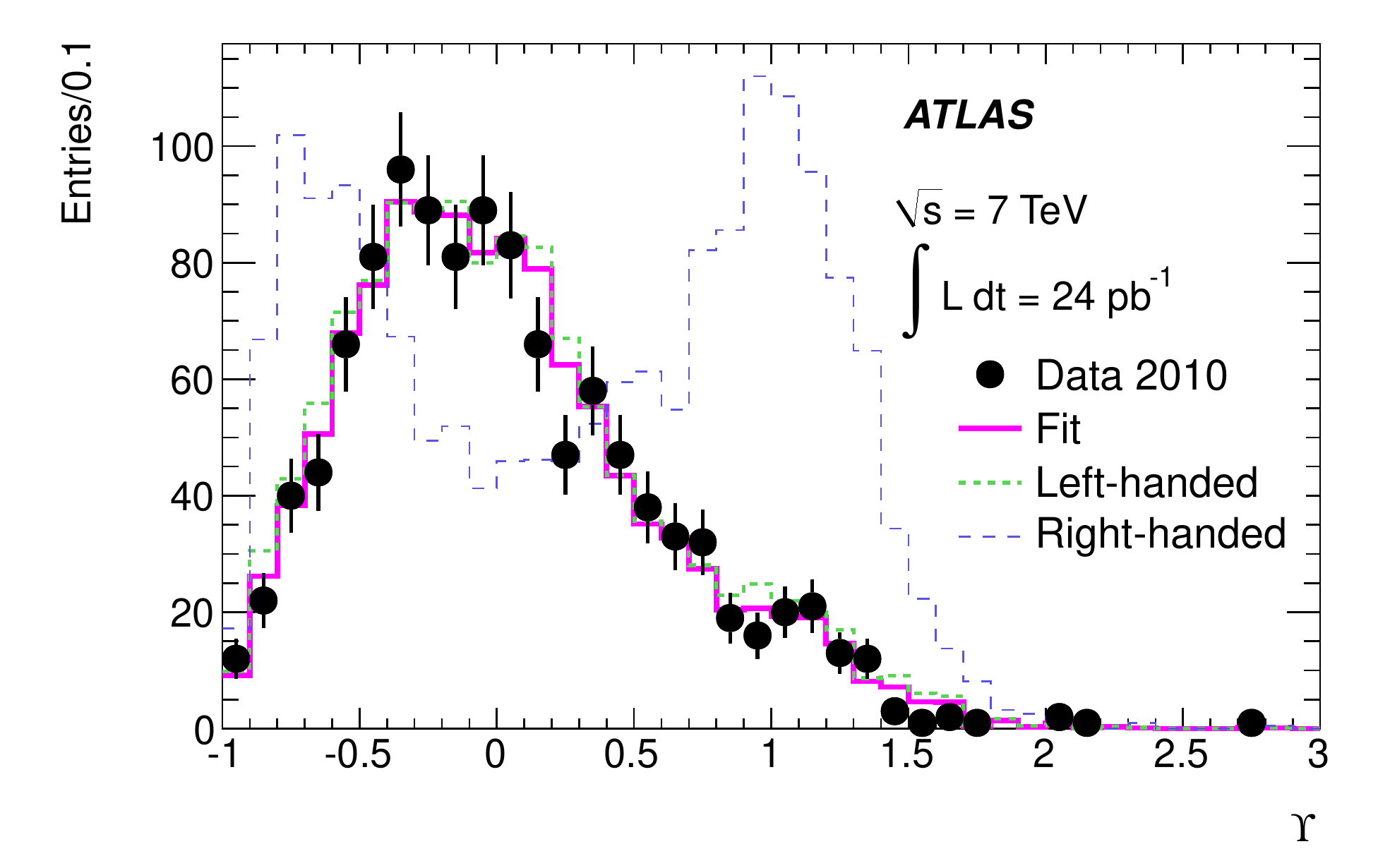}
\caption{Simulated signal and background templates for left-handed and right-handed $\tau$~decays along with the observed charged asymmetry distribution in data. The best fit resulting from maximizing the likelihood is plotted in bold.}
\label{fig:fit}      
\end{figure}

The fit is performed over the range $-1 \leq \Upsilon \leq 3$ with bins of width $\Delta\Upsilon = 0.1$. Figure~\ref{fig:fit} shows the left-handed and right-handed templates plotted together with the 
observed charged asymmetry distribution in data along with the resulting fit. The fitted value of the $\tau$~polarization and its associated statistical uncertainty is \linebreak $\Ptau = -1.06 \pm 0.04(stat)$. 
As an assessment of the quality of the fit, the $\chi^2$ per degree of freedom is calculated using only the statistical uncertainties on the data sample and with the bins 
in the range $1.5<\Upsilon<3.0$ merged due to the low number of events in this region. With 22 degrees of freedom ($ndf$) the resulting value is $\chi^2/ndf~=~$1.1. The value of the Monte Carlo normalization parameter is given by $N_{\mathrm{MC}} = 0.98 \pm 0.04(stat)$. 

\section{Systematic Uncertainties} 
\label{sec:Systematics}
This analysis relies on the prediction of the shapes of the 
left-handed and the right-handed templates, which include the simulated signal
and backgrounds. Systematic uncertainties are evaluated for their effect on the shape of the $\Upsilon$ distribution, as well as for any changes in the relative acceptance of the signal and background events. 

For each source of systematic uncertainty, new templates are constructed and fit to the data. The corresponding uncertainty on \Ptau~is
taken as the difference between the fit values obtained with the nominal and the new templates.
The total systematic uncertainty is calculated as the sum in quadrature of the individual uncertainties. The results are presented in Table~\ref{tab:Systsummary}
and the various sources of systematic uncertainty are discussed below.

\paragraph{Energy Scale and Resolution:}

The dominant source of systematic uncertainty arises from the calibration of energy scales 
used to make measurements of the $\tau$~candidate and cluster $\pt$, and the event \met~ 
and \metsig. The cluster energy scale uncertainty varies between 10\% for low~\pt~and 3\% for high~\pt~clusters in the central region of the detector defined by $|\eta|~<~3.2$ and is estimated to be 10\% in the forward region $|\eta|~>~3.2$. The $\tau$~energy scale uncertainty similarly varies with $\tau$~lepton~$\pt$ and $\eta$ and ranges between 2.5\% and 10\%~\cite{BDT1}. 

The systematic uncertainty attributed to the energy scales in the extraction of $\Ptau$ is assessed separately for the central and forward regions of the detector. In the central region, the cluster energy scale is varied for all clusters with $|\eta|~<~3.2$ with a $\pt$-dependent scale factor, and $\met$ and $\metsig$ are recalculated. The $\tau$~energy scale is simultaneously varied with $\tau$~lepton~$\pt$ and $\eta$, and new Monte Carlo templates are generated. To assess the effect of the cluster energy calibration in the forward region on the $\met$ and $\metsig$, the cluster energy scale is varied by its uncertainty, 10\%, for all clusters reconstructed with $|\eta|~>~3.2$.

The resolution of \met~ components is given by $\alpha\sqrt{\sum \et}$.
The scaling factor $\alpha$ was measured using minimum bias events to be 0.49$\sqrt{\GeV}$~\cite{ETMISS}.
In the presence of high~\pt~jets the resolution is degraded and Monte Carlo simulations yield $\alpha~=~0.55\sqrt{\GeV}$. The sensitivity to the acceptance difference for signal and various backgrounds due to the 
resolution of \met~is evaluated by adding a Gaussian term on the $x$ and $y$ components of the \met~in order
to reproduce the \met~resolution in the presence of high~\pt~jets. 
In addition, the effect due to an imperfect modeling of the response of the inner ring cells of the forward calorimeter (FCal)
($4.5~<~|\eta|~<~4.9$) is evaluated by measuring the impact of excluding these cells from the computation of \met~and \metsig.

The track momentum resolution was studied using \Zmumu~decays and shown to be well modeled in the Monte Carlo simulations.
Any associated systematic uncertainty has a negligible impact on the $\tau$~polarization measurement.

\paragraph{Physics Object Identification and Trigger:}

The uncertainty on the identification and reconstruction efficiencies for $\tau$~candidates
originates from different simulation conditions such
as underlying event models, detector geometry, hadronic
shower modeling and noise thresholds for calorimeter cells used to build the
identification observables. It varies as a function of the $\tau$~lepton~\pt~and the multiplicity of primary vertices in the event, between 5\% and 13\%~\cite{WtauXS}.
To assess the impact on the templates, these uncertainties are applied to
the simulated left-handed and the right-handed $\Wtaunu$ signal as well as to the \Ztautau~background samples.
Other sources of systematic uncertainty are found to have a negligible impact. These include possible mis-modeling of: the jet fake rate in the $W$+jets events~\cite{WtauXS}; the electron veto in the $\tau$~reconstruction; and lepton selection efficiencies. 

The systematic uncertainty on the efficiency of the combined $\tau$~and \met~trigger
is derived from the level of agreement between the measured trigger responses of the two
trigger components in data and simulation. The correlation between
the two trigger chains is found to be much smaller than their respective errors and is neglected.
The systematic uncertainty for $\tau$~lepton~\pt~between 20 and 30~GeV (turn-on region), where modeling the trigger response is the most challenging, 
is taken to be twice the difference between the measured and the simulated values.
The total systematic uncertainty after the combination of the different trigger components is 14\%
for $\tau$~candidates in the turn-on region and 4\% for $\tau$~candidates with \pt~between 30 and 60~GeV.
The impact on the templates is evaluated by scaling the numbers of Monte Carlo events
by the above uncertainties.
\paragraph{Monte Carlo Modeling and Normalization:}

The uncertainty due to Monte Carlo modeling of the signal is evaluated via a comparison
of the left-handed HERWIG++ signal sample
with an alternative left-handed $\Wtaunu$ signal sample generated with PYTHIA with $\tau$~leptons decayed by TAUOLA.
The main difference concerns the acceptance for the $\met{}$ significance cut which is found 
to originate from differences in modeling of the underlying event. 
The total number of signal and background events estimated using PYTHIA is found to be 18\% lower than when using HERWIG++.
Left-handed and right-handed signal templates were prepared by 
splitting the PYTHIA sample into two sub-samples of comparable size, one of which was
re-weighted event-wise such that the resulting distribution of $\Upsilon$ emulates that of right-handed $\tau$~leptons, using the TauSpinner tool~\cite{TauSpinnerAlg1,TauSpinERWZW}.
The remaining left-handed sub-sample was left unaltered.

The combined statistical, systematic and acceptance uncertainties on the measured $W$, $Z$ and \ttbar~cross-sections
are used to evaluate the sensitivity of the $\tau$~polarization measurement to the relative acceptance differences in the signal and backgrounds
in the templates. The systematic uncertainty arising from the \ttbar~cross-section uncertainty was found to be negligible. 
The uncertainty on the integrated luminosity is removed by determining the $N_{\mathrm{MC}}$ and $N_{\mathrm{MJ}}$ in the fit.
The treatment of pile-up effects in the Monte Carlo
samples is found to have a negligible impact on the $\tau$~polarization measurement.

\begin{table}[h]
\begin{center}
\caption{Summary of the systematic uncertainty on \Ptau.}
\label{tab:Systsummary}
\begin{tabular}{lcc}
\hline\noalign{\smallskip}
Source & +$\Delta$\Ptau & -$\Delta$\Ptau\\
\noalign{\smallskip}\hline\noalign{\smallskip}
Energy scale central & 0.042 & 0.063\\
Energy scale forward & 0.007 &  0.002\\
$\met{}$ resolution & 0.014 & -\\
No FCal & 0.003 &-\\
\hline
$\tau$~identification& 0.005 & 0.006\\
Trigger & 0.007 & 0.006\\
\hline
MC model & 0.020 & 0.020\\
$W$ cross-section& 0.005 & 0.005\\
$Z$ cross-section& 0.006 & 0.006\\
\noalign{\smallskip}\hline\noalign{\smallskip}
Combined & 0.05 & 0.07\\
\noalign{\smallskip}\hline\noalign{\smallskip}
\end{tabular}
\end{center}
\end{table}
\section{Results}
\label{sec:Results}

The result of the $\tau$~polarization measurement in the selected sample of \Wtaunu~decays is:
\[ \Ptau = -1.06\; \pm 0.04\; (stat)\; ^{+0.05}_{-0.07}\;(syst). \]
The central value of \Ptau~falls outside the physically allowed
range of [--1,1].  A Bayesian approach is used to determine a 95\% credibility interval from the posterior probability distribution. A uniform prior was assumed in the interval [--1, 1] and the likelihood was approximated by a normal distribution with a mean of --1.06 and a width given by the upper limit of the combined statistical and systematic uncertainties. \Ptau~is found to lie within the 95\% credibility interval [--1, --0.91].

\section{Summary and Conclusions}
\label{sec:summary}

The $\tau$ polarization has been measured in the single-track hadronic decays of $\tau$~leptons in $\Wtaunu$ events with data collected in 2010 with the ATLAS detector corresponding to an integrated luminosity of 24~\ipb. The measurement was carried out in the phase space specified in Table~\ref{tab:truespace} and 
the measured value is consistent with the Standard Model prediction within the statistical and systematic uncertainties. This result marks the first measurement of $\tau$~polarization in a hadron collider and the first such measurement with $\tau$~leptons in $W$ boson decays.  The efficacy of the method for extracting $\tau$~polarization and the relatively small systematic uncertainties associated with this measurement confirm the potential of successful future applications of this technique.

\section{Acknowledgements}
\label{sec:Ack}

We thank CERN for the very successful operation of the LHC, as well as the
support staff from our institutions without whom ATLAS could not be
operated efficiently.

We acknowledge the support of ANPCyT, Argentina; YerPhI, Armenia; ARC,
Australia; BMWF, Austria; ANAS, Azerbaijan; SSTC, Belarus; CNPq and FAPESP,
Brazil; NSERC, NRC and CFI, Canada; CERN; CONICYT, Chile; CAS, MOST and NSFC,
China; COLCIENCIAS, Colombia; MSMT CR, MPO CR and VSC CR, Czech Republic;
DNRF, DNSRC and Lundbeck Foundation, Denmark; EPLANET and ERC, European Union;
IN2P3-CNRS, CEA-DSM/IRFU, France; GNAS, Georgia; BMBF, DFG, HGF, MPG and AvH
Foundation, Germany; GSRT, Greece; ISF, MINERVA, GIF, DIP and Benoziyo Center,
Israel; INFN, Italy; MEXT and JSPS, Japan; CNRST, Morocco; FOM and NWO,
Netherlands; RCN, Norway; MNiSW, Poland; GRICES and FCT, Portugal; MERYS
(MECTS), Romania; MES of Russia and ROSATOM, Russian Federation; JINR; MSTD,
Serbia; MSSR, Slovakia; ARRS and MVZT, Slovenia; DST/NRF, South Africa;
MICINN, Spain; SRC and Wallenberg Foundation, Sweden; SER, SNSF and Cantons of
Bern and Geneva, Switzerland; NSC, Taiwan; TAEK, Turkey; STFC, the Royal
Society and Leverhulme Trust, United Kingdom; DOE and NSF, United States of
America.

The crucial computing support from all WLCG partners is acknowledged
gratefully, in particular from CERN and the ATLAS Tier-1 facilities at
TRIUMF (Canada), NDGF (Denmark, Norway, Sweden), CC-IN2P3 (France),
KIT/GridKA (Germany), INFN-CNAF (Italy), NL-T1 (Netherlands), PIC (Spain),
ASGC (Taiwan), RAL (UK) and BNL (USA) and in the Tier-2 facilities
worldwide.

\bibliographystyle{spphys} 
\bibliography{TauPolar}

\begin{thebibliography}{10}
\providecommand{\url}[1]{{#1}}
\providecommand{\urlprefix}{URL }
\expandafter\ifx\csname urlstyle\endcsname\relax
  \providecommand{\doi}[1]{DOI \discretionary{}{}{}#1}\else
  \providecommand{\doi}{DOI \discretionary{}{}{}\begingroup
  \urlstyle{rm}\Url}\fi

\bibitem{HIGGSATLAS}
{ATLAS Collaboration}, Phys. Lett. B \textbf{710}, 49 (2012).

\bibitem{HIGGSCMS}
{CMS Collaboration}, Phys. Lett. B. \textbf{710}, 26 (2012).

\bibitem{MSSMTRY}
{S. Dittmaier et al.}, arXiv:1101.0593  (2011).

\bibitem{PDG}
{The Particle Data Group, K. Nakamura et al.}, J. Phys. G \textbf{37}, 260
  (2010).

\bibitem{B235}
{K. Hagiwara, A.D. Martin, D. Zeppenfeld}, Phys. Lett. B \textbf{235}, 198
  (1990).

\bibitem{C48}
A.~Rouge, Z. Phys. C \textbf{48}, 75 (1990).

\bibitem{ATLASDet}
{ATLAS Collaboration}, JINST 3 \textbf{S08003} (2008).

\bibitem{TAUTRIG}
{ATLAS Collaboration}, Eur. Phys. J. C \textbf{72}, 1849 (2012).

\bibitem{HERWIGPP}
{M. Bahr et al.}, Eur. Phys. J C \textbf{58}, 639 (2008).

\bibitem{PDF}
{A. Sherstnev, R.S. Thorne}, Eur. Phys. J. C Part. Fields \textbf{55}, 553
  (2008).

\bibitem{UEHER}
{S. Gieseke et al.}, arXiv:1102.1672  (2011).

\bibitem{PYTHIA}
{T. Sjostrand}, Comput. Phys. Commun. \textbf{82}, 74 (1994).

\bibitem{MCATNLO}
{S. Frixione, B.R. Webber}, J. High Energy Phys. \textbf{0206}, 029 (2002).

\bibitem{HERWIG}
{G. Corcella et al.}, J. High Energy Phys. \textbf{0101} (2001).

\bibitem{JIMMY}
{J.M. Butterworth, J.R. Forshaw, M.H. Seymour}, Z. Phys. C \textbf{72}, 637
  (1996).

\bibitem{TAUOLA1}
{S. Jadach et al.}, Comput. Phys. Commun. \textbf{64}, 275 (1990).

\bibitem{TAUOLA2}
M.~Jezabek, et~al., Comput. Phys. Commun. \textbf{70}, 69 (1992).

\bibitem{TAUOLA3}
R.~Decker, et~al., Comput. Phys. Commun. \textbf{76}, 361 (1993).

\bibitem{PHOTOS}
{E. Barberio, B.V. Eijk, Z. Was}, Comput. Phys. Comm. \textbf{66}, 115 (1991).

\bibitem{GEANT4A}
{ATLAS Collaboration}, Eur. Phys. J. C \textbf{70}, 82 (2010).

\bibitem{GEANT4}
{GEANT4 Collaboration}, Nucl. Inst. Meth. A \textbf{506}, 250 (2003).

\bibitem{ATLASMC10}
{ATLAS Collaboration}, {ATLAS-PHYS-PUB-2010-014}  (2010).
\newblock \url{https://cdsweb.cern.ch/record/1303025}

\bibitem{WtauXS}
{ATLAS Collaboration}, Phys. Lett. B \textbf{706}, 276 (2012).

\bibitem{ELERECO}
{ATLAS Collaboration}, Eur. Phys. J. C \textbf{72}, 1909 (2012).

\bibitem{ANTIKT}
{M. Cacciari, G.P. Salam, G. Soyez}, J. High Energy Phys. \textbf{0804}, 063
  (2008).

\bibitem{JES}
{ATLAS Collaboration}, Eur. Phys. J. C \textbf{71}, 1512 (2011).

\bibitem{JCAL}
{T. Barillari, et al.}, {ATL-LARG-PUB-2009-001}  (2009).
\newblock \url{https://cdsweb.cern.ch/record/1112035}

\bibitem{TOPO}
{W. Lampl, et al.}, {ATL-LARG-PUB-2008-002}  (2008).
\newblock \url{https://cdsweb.cern.ch/record/1099735}

\bibitem{BDT1}
{ATLAS Collaboration}, ATLAS-CONF-2011-077  (2011).
\newblock \url{https://cdsweb.cern.ch/record/1353226}

\bibitem{ETMISS}
{ATLAS Collaboration}, Eur. Phys. J. C \textbf{72}, 1844 (2012).

\bibitem{WZxsec}
{ATLAS Collaboration}, Phys. Rev. D \textbf{72}, 072004 (2012).

\bibitem{ttbar}
{ATLAS Collaboration}, Phys. Lett. B \textbf{711}, 244 (2012).

\bibitem{B306}
{M. Davier, L. Duflot, F. Le Diberder, A. Rouge}, Phys. Lett. B \textbf{306},
  411 (1993).

\bibitem{TauSpinnerAlg1}
{Z. Czyczula, T. Przedzinski, Z. Was}, Eur. Phys. J. C \textbf{72}, 1988
  (2012).

\bibitem{TauSpinERWZW}
{T. Pierzchala et al.}, Acta Phys. Pol. B \textbf{32}, 1277 (2001).

\end{thebibliography}

\newpage
\clearpage
\onecolumn
\begin{flushleft}
{\Large The ATLAS Collaboration}

\bigskip

G.~Aad$^{\rm 48}$,
B.~Abbott$^{\rm 111}$,
J.~Abdallah$^{\rm 11}$,
S.~Abdel~Khalek$^{\rm 115}$,
A.A.~Abdelalim$^{\rm 49}$,
A.~Abdesselam$^{\rm 118}$,
O.~Abdinov$^{\rm 10}$,
B.~Abi$^{\rm 112}$,
M.~Abolins$^{\rm 88}$,
O.S.~AbouZeid$^{\rm 158}$
% ...
\bigskip

% ...
$^{10}$ Institute of Physics, Azerbaijan Academy of Sciences, Baku, Azerbaijan\\
$^{11}$ Institut de F\'isica d'Altes Energies and Departament de F\'isica de la Universitat Aut\`onoma  de Barcelona and ICREA, Barcelona, Spain\\
% ...
$^{48}$ Fakult\"{a}t f\"{u}r Mathematik und Physik, Albert-Ludwigs-Universit\"{a}t, Freiburg i.Br., Germany\\
$^{49}$ Section de Physique, Universit\'e de Gen\`eve, Geneva, Switzerland\\
% ...
$^{88}$ Department of Physics and Astronomy, Michigan State University, East Lansing MI, United States of America\\
% ...
$^{111}$ Homer L. Dodge Department of Physics and Astronomy, University of Oklahoma, Norman OK, United States of America\\
$^{112}$ Department of Physics, Oklahoma State University, Stillwater OK, United States of America\\
% ...
$^{115}$ LAL, Univ. Paris-Sud and CNRS/IN2P3, Orsay, France\\
% ...
$^{118}$ Department of Physics, Oxford University, Oxford, United Kingdom\\
% ...
$^{158}$ Department of Physics, University of Toronto, Toronto ON, Canada\\
% ...
\end{flushleft}


\end{document}